\documentclass[%
twocolumn,
superscriptaddress,
floatfix,
amsmath,amssymb,
aps,
pre,
a4paper,
]{revtex4-2}

\usepackage{newtxmath}
\usepackage{blindtext}
\usepackage{graphicx}% Include figure files
\usepackage{dcolumn}% Align table columns on decimal point
\usepackage{bm}% bold math
\usepackage{lipsum}

\usepackage[hidelinks,unicode=true]{hyperref}
\hypersetup{colorlinks=true,
  linkcolor=blue,
  urlcolor=blue,
  citecolor=blue,
  pdfhighlight=/N
}

\usepackage{amsmath}

\usepackage{siunitx}
\usepackage{soul}
\usepackage{xcolor}
\setstcolor{red}

\begin{document}

\title{Phase synchronization of fish schools through spatial gaps}

\author{Elena G. de Lamo}

\affiliation{Departament de Física, Universitat Politècnica de Catalunya, Campus
  Nord B4, 08034 Barcelona, Spain}

\author{Óscar Sánchez}
\affiliation{Departament de Física, Universitat Politècnica de Catalunya, Campus
  Nord B4, 08034 Barcelona, Spain}

\author{M. Carmen Miguel}

\affiliation{Departament de Física de la Matèria Condensada, Universitat de
  Barcelona, Martí i Franquès 1, 08028 Barcelona, Spain}

\affiliation{Institute of Complex Systems (UBICS), Universitat de Barcelona,
  08028 Barcelona, Spain}

\author{Romualdo Pastor-Satorras}
\email{romualdo.pastor@upc.edu}
\affiliation{Departament de Física, Universitat Politècnica de Catalunya, Campus
  Nord B4, 08034 Barcelona, Spain}

\begin{abstract}
We investigate how sensory coupling mediates phase coherence between spatially
separated, confined active systems. Using two fish schools separated by a
water-filled transparent gap, we isolate vision as the primary interaction channel. Under boundary confinement, each school's center-of-mass heading exhibits noisy bistable dynamics, spontaneously switching between two antiparallel directions along the partition wall. Turning events in one school bias the orientation of the other, driving distance-dependent phase synchronization. We model this behavior using two coupled stochastic bistable oscillators whose exact solution quantitatively reproduces the empirical results. Our work establishes a precise connection between nonequilibrium active systems and coupled stochastic oscillators, offering insight into boundary-spanning information transfer in living collectives.
\end{abstract}

\maketitle

\section{Introduction}

The emergence of macroscopically coordinated states from local, non-equilibrium interactions is a defining feature of active matter systems~\cite{marchettiHydrodynamicsSoftActive2013}. In a biological setting, collective animal behavior emerges from information exchanged among individuals. Directional changes in bird flocks propagate as traveling waves or diffuse across the group~\cite{cavagnaScalefreeCorrelationsStarling2010,cavagnaPhysicsFlockingCorrelation2018,papadopoulouDiffusionCollectiveTurns2023}, whereas in fish schools collective turns frequently take the form of cascade-like events initiated by small subsets of individuals~\cite{puySelfsimilarityTurningAvalanches2023,mugicaScalefreeBehavioralCascades2022}.
While the microscopic rules governing intra-group interactions—--ranging from topological metrics to selective attention—--have been extensively studied~\cite{balleriniInteractionRulingAnimal2008,cavagnaBirdFlocksCondensed2014,katzInferringStructureDynamics2011,puySelectiveSocialInteractions2024}, much less is known about the effective interactions and information transfer that operate across physical boundaries between spatially segregated subgroups.

In the wild, many animal collectives display fission–fusion dynamics, repeatedly splitting and merging in response to predation risk, habitat heterogeneity, or other ecological pressures~\cite{https://doi.org/10.1111/j.1600-0706.2011.19685.x,JACOBS2010678,kelleyPredationRiskShapes2011,ventejouBehavioralTransitionFish2024}. Even when physically separated by environmental gaps, subgroups may remain within sensory range, exploiting long-range sensory channels—--such as vision, acoustics, and chemical signaling--—that extend beyond immediate physical proximity. However, because traditional models and theoretical approaches to collective motion usually assume local, short-range alignment rules, they provide limited insight into communication across gaps and inherently fail to capture information transfer across physical voids. This leaves a fundamental question largely unresolved: Whether coordinated information transfer can persist between physically separated subgroups lacking direct contact, and how non-contact communication channels shape their collective dynamics.

Here, we address this question experimentally by studying information transfer between two spatially separated fish schools that interact through vision. Using a partitioned arena, we impose physical separation that suppresses hydrodynamic and contact-mediated coupling while preserving visual access, thereby isolating the role and spatial range of sensory mediation in group coordination. Because vision has been identified as a dominant modality in schooling \cite{SchoolingFishesSchool,xueTuningSocialInteractions2023}, it provides a natural primary communication channel for our setup.

A powerful framework for probing these dynamics is boundary-induced confinement. In active systems, physical boundaries can break spatial symmetry, generate steady-state currents, and induce sharp phase transitions~\cite{RevModPhys.96.031003,Galajda2007,Bechinger2016}. In our experimental geometry, compartment  boundaries restrict the collective motion of each school to a dominant spatial dimension. This dimensional reduction enables a low-dimensional coarse-grained description in which each school behaves as a noisy, bistable unit, stochastically switching between two preferred, antiparallel directions—--a classic form of stochastic bistability ubiquitous in physical and biological systems alike~\cite{pikovsky2001synchronization,riskenFokkerPlanckEquationMethods1996}. Importantly, this bistability is not an assumption imposed a priori, but emerges directly from the empirical center-of-mass trajectories of the schools.

To explain these findings, we formulate a minimal stochastic model of two coupled noisy bistable oscillators, with each oscillator representing the collective orientation of one school and a distance-dependent coupling strength quantifying visual information transfer between groups. The model admits an exact stationary solution that quantitatively reproduces the experimentally observed distance-dependent phase synchronization.

Beyond capturing collective dynamics, our model establishes a predictive quantitative framework for information transmission in spatially extended biological systems. Fitting the model to empirical data reveals that effective visual coupling decays with intergroup separation,  exhibiting a faster-than-exponential relaxation at short distances followed by an exponential decay at longer distances, consistent with simple visual attenuation. This finding directly connects collective motion to sensory ecology, where spatial signal detection and information processing are central to behavior \cite{stevensSensoryEcologyBehaviour2013,EcologyVisionLythgoe}.

\section{Experiment}

To test these ideas, we monitor schools of black neon tetra (\textit{Hyphessobrycon herbertaxelrodi}) inside a quasi-two-dimensional square tank partitioned by two impermeable, transparent walls into three sections. This geometry creates two outer rectangular compartments, each housing one school, separated by a central water-filled gap of width $d$ that acts as a tunable visual bottleneck controlling information exchange between groups. Crucially, the solid partitions eliminate hydrodynamic coupling--—preventing fluid momentum transfer and suppressing pressure-mediated disturbances—--thereby isolating visual access as the dominant channel for intergroup interaction. Boundary confinement within the rectangular compartments further alters collective motion relative to unbounded conditions; such geometric constraints are known to strongly influence group dynamics and promote structured states such as milling formation \cite{caloviDisentanglingModelingInteractions2018, lafouxConfinementdrivenStateTransition2024}. In Supplemental Video~\ref{vid:video1} we show a rendering of one of our experiments. Details of the experimental setup are provided in Appendix~\ref{sec:experimental_methods}.

We analyze the macroscopic dynamics of each school by its center-of-mass trajectory, defined as $\vec{R}(t) = \sum_{i=1}^N\vec{r}_i(t) /N$, where $\vec{r}_i(t)$ is the position of fish $i$ at time $t$ and $N$ is the group size. The trajectories reveal predominantly back-and-forth motion along the long axis of each compartment, as evidenced by the two-dimensional spatial probability distributions $P(X,Y)$ of the center-of-mass position, see  Supplemental Figure (SF) Fig.~\ref{fig:spatial_distribution}. This directional bias reflects the combined effects of visually induced accumulation near the partition, wall alignment \cite{caloviDisentanglingModelingInteractions2018}, and the longer uninterrupted swimming path imposed by the rectangular geometry. Consequently, the collective dynamics can be reduced to an effectively one-dimensional description along the compartment's major axis.

\begin{figure}[t]
    \centering
    \includegraphics[width=0.9\linewidth]{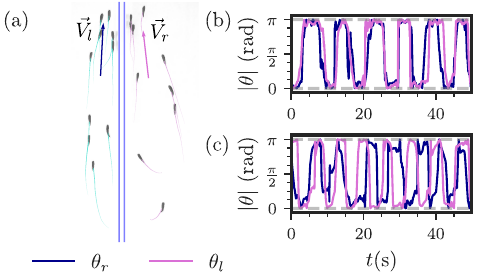}
    \caption{(a) Representative frame of the experimental arena: a square tank partitioned by two transparent walls (blue dashed lines) separated by a water-filled gap of $d=4$cm. Tracked individual trajectories for the left and right schools are overlaid; arrows indicate the center‑of‑mass velocity vectors, plotted at each group’s center of mass. (b, c) Time series of the center‑of‑mass orientation $\theta$ for both schools at $d=4$cm (b) and with an opaque wall (c), showing bistable dynamics with preferred directions at $\theta=0$ and $|\theta|=\pi$ (grey dashed lines). At short separation the two traces are strongly synchronized, but synchronization is lost for the opaque wall. }
    \label{fig:1}
\end{figure}

This dimensional reduction allows us to characterize the collective dynamics using a single macroscopic coordinate: the orientation angle $\theta = \angle(\hat{y}, \vec{V})$ of the school's center-of-mass velocity $\vec{V}$, measured relative to the unit vector $\hat{y}$ along the compartment's long axis. Figure~\ref{fig:1}(a) shows a representative experimental snapshot at an intermediate gap $d = 4$cm, with tracked individual trajectories and the aggregate velocity vectors $\vec{V}_l$ and $\vec{V}_r$ of the left and right schools. In this confined geometry, the collective velocity breaks directional symmetry along the boundaries, and each school spontaneously switches between two antiparallel preferred states parallel to the partition walls.

Fig.~\ref{fig:1}(b) shows the temporal evolution of the center-of-mass orientation for the left, $\theta_l$, and right, $\theta_r$, schools. Each trace exhibits spontaneous switches between the two preferred directions.
At short gap distances ($d = 4\text{cm}$), the orientation trajectories $\theta_r(t)$ and $\theta_l(t)$ are visibly synchronized, see Fig.~\ref{fig:1}(b), signaling an effective visual coupling between the two schools. Conversely, when visual communication is completely blocked by an opaque partition, functionally equivalent to an infinite separation limit ($d \to \infty$), the trajectories decorrelate, exhibiting independent switching, see Fig.~\ref{fig:1}(c). Inspection of trajectories across intermediate gaps confirms that this synchronization weakens systematically as separation increases (see Fig.~\ref{fig:bistability}). Together, these features—--individual bistable switching combined with distance-dependent phase alignment—--naturally motivate a coarse-grained model of the system as two noisy bistable oscillators governed by a separation-dependent coupling strength. Therefore, this coarse-grained orientational dynamics admit an effective equilibrium description even though the underlying microscopic system is manifestly out of equilibrium.

\section{Model}

Motivated by the experimental observations, we describe the collective orientations of the right ($\theta_r$) and left ($\theta_l$) schools by two coupled noisy overdamped bistable oscillators
\begin{equation}
    \begin{aligned}
        \dot{\theta}_l &= -2K\sin(2\theta_l) +  J\sin(\theta_r - \theta_l) + \sqrt{2D}\,\xi_l, \\
        \dot{\theta}_r &= -2K\sin(2\theta_r) +  J\sin(\theta_l - \theta_r) +
        \sqrt{2D}\,\xi_r.
    \end{aligned}
    \label{eq:thetas}
\end{equation}
The single-school dynamics are governed by a double-well potential $U(\theta) = -K \cos(2\theta)$, whose barrier height $K$ enforces the observed bistability at $\theta = 0$ and $\theta = \pm\pi$. Visual synchronization is mediated by a Kuramoto-type coupling of strength $J \ge 0$, which favors parallel alignment between collective headings \cite{kuramotoSelfentrainmentPopulationCoupled1975}. Behavioral indecision, confinement, and finite-size fluctuations  are modeled as Gaussian white noise $\xi_i(t)$, $i = r, l$, with amplitude $D$, characterized by zero mean $\langle\xi_i(t)\rangle = 0$ and delta-correlations $\langle\xi_i(t)\xi_j(t')\rangle = \delta_{ij}\delta(t-t')$.

This model satisfies detailed balance and its stationary distribution can be obtained analytically. Instead of solving Eq.~\eqref{eq:thetas} directly, we transform the individual school headings $(\theta_r,\theta_l)$ into relative and total orientation coordinates, $\varphi = \theta_r - \theta_l$ and $\psi = \theta_r + \theta_l$, respectively. The phase difference $\varphi$ acts as a macroscopic synchronization order parameter: $\varphi = 0$ denotes parallel alignment, whereas $\varphi = \pm\pi$ denotes antiparallel alignment. In these variables, the stochastic equations become
\begin{equation}
    \begin{aligned}
    \dot{\varphi} &= (2J-4K\cos\psi)\sin\varphi +\sqrt{4D}\xi_\varphi,\\
    \dot{\psi} &= -4K\sin\psi\cos\varphi +\sqrt{4D}\xi_\psi,
    \end{aligned}
\end{equation}
with $\xi_\varphi = \xi_r - \xi_l$ and $\xi_\psi = \xi_r + \xi_l$.
The joint stationary distribution  can be obtained from the corresponding Fokker-Planck equation using standard tools from stochastic processes~\cite{gardinerHandbookStochasticMethods1985}, and it takes the form
\begin{equation}
    P(\varphi,\psi) = \frac{1}{ Z_1} \exp\left[  \lambda\cos\varphi + 2\kappa \cos\psi \cos\varphi\right],
    \label{eq:joint}
\end{equation}
while the corresponding marginal distribution of the phase difference $\varphi$ is
\begin{equation}
    P(\varphi)=\int P(\varphi,\psi) \;d\psi = \frac{1}{Z_2}e^{\lambda\cos\varphi}I_0(2\kappa\cos\varphi),
    \label{eq:analyticalsolution}
\end{equation}
where $I_0(z)$ is the modified Bessel function of the first kind of order zero~\cite{abramovitz1972}, $Z_1$ and $Z_2$ are a normalization contants, and $\lambda = J/D$ and $\kappa = K/D$ denote the effective coupling and bistability strength, respectively.

\begin{figure}[t]
    \centering
    \includegraphics[width=0.9\linewidth]{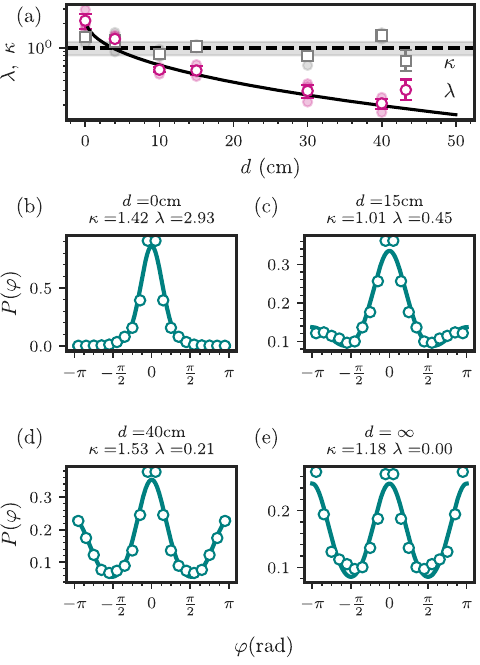}
    \caption{(a) Effective coupling parameter $\lambda$ and bistability parameter $\kappa$ (log scale) inferred by fitting the empirical phase-difference distributions as a function of gap distance. Solid markers: mean $\lambda$ and $\kappa$ values across replicates; error bars: standard deviation. Semi‑transparent markers: individual‑replicate estimates. The black dashed line and shaded band denote the mean and standard deviation of $\kappa$ at $d=\infty$. The black curve is a stretched‑exponential fit to the mean values of $\lambda$. (b - e) Empirical distributions of the phase difference $\varphi=\theta_r-\theta_l$ for selected gap distances. The distributions evolve from a single peak at $\varphi=0$ at small $d$, indicating strong synchrony, to symmetric peaks at $\varphi=0,\pm\pi$ for the opaque partition, $d\to\infty$, corresponding to uncoupled bistable switching. Distributions were symmetrized under $\varphi\mapsto-\varphi$ to enforce the expected parity. Points denote experimental data, and solid lines are fits to the analytical stationary solution, Eq.\eqref{eq:analyticalsolution}.  }
    \label{fig:2}
\end{figure}

\section{Model fitting and experimental validation}

To quantitatively map the empirical trajectories onto the stochastic model, we infer the dimensionless parameters---the intragroup bistability strength $\kappa$ and the intergroup visual coupling  $\lambda$---by unbinned maximum-likelihood estimation (MLE)~\cite{gregory2010bayesian} using the analytical stationary phase-difference density $P(\varphi)$ in Eq.~\eqref{eq:analyticalsolution}. Complete likelihood formulations and numerical calibration are provided in Appendix~\ref{sec:mle_fitting}.

In the absence of visual communication, realized experimentally by an opaque partition and corresponding to $d\to\infty$, intergroup coupling should vanish, yielding $\lambda = 0$. In this limit, the theoretical distribution reduces to $P(\varphi) \propto I_0(2\kappa \cos\varphi)$, with two equal-height peaks at $\varphi = 0$ (parallel alignment) and $\varphi = \pm\pi$ (anti-parallel alignment), reflecting independent, uncoordinated switching, see Fig.~\ref{fig:2}(e). Fitting this uncoupled baseline across experimental replicates yields an average bistability parameter $\kappa \approx 1.0\pm 0.2$ and a residual coupling $\lambda \approx 0.03\pm 0.04$, consistent with zero within statistical uncertainty. For finite separations, the inferred $\kappa$ remains approximately constant across all tested gap distances, Fig.~\ref{fig:2}(a), with an average value $\kappa \approx 1.1\pm 0.2$, confirming that the intrinsic single-school bistable switching mechanism is robust and independent of partition geometry.

When visual interaction is permitted across a gap of width $d$, the full two-parameter density, $P(\varphi) \propto e^{\lambda \cos\varphi} I_0(2\kappa \cos\varphi)$, quantitatively captures the empirical phase-difference distributions across all conditions, see Fig.~\ref{fig:2}(b-e) and Fig.~\ref{fig:phi}. As $d$ decreases and visual interaction strengthens, $P(\varphi)$ exhibits a qualitative structural crossover: the secondary peak at $\varphi = \pm\pi$ (anti-parallel alignment) is progressively suppressed, while a dominant central peak emerges at $\varphi = 0$ (parallel synchronization) [Fig.~\ref{fig:2}(b),(e)]. In the limit $d = 0$, where the two schools are separated only by a transparent wall, the anti-parallel state disappears entirely, yielding a single sharp peak at $\varphi = 0$ that indicates near-perfect synchronization limited only by finite-size behavioral fluctuations, Fig.~\ref{fig:2}(b)

The inferred coupling parameter $\lambda(d)$ decreases systematically with gap distance $d$. This decline reflects the attenuation of visual communication as larger separations reduce both visual angle and perceptual contrast. This trend is consistent with recent evidence that synchronization in schooling fish is light- and vision-dependent and degrades when perceptual cues are reduced \cite{xueTuningSocialInteractions2023}.
Quantitatively, although classical models of aquatic visual detection predict simple exponential attenuation of optical contrast with distance \cite{aksnesRevisedModelVisual1997}, the inferred coupling strength is better described over the tested separations by a stretched exponential:
\begin{equation}
\lambda(d) = \lambda_0 \exp\!\left[ -\left(d/\ell \right)^\beta \right],
\label{eq:stretched_exp}
\end{equation}
where $\ell$ is the characteristic visual interaction range and $\beta < 1$ the stretching exponent, see Fig.~\ref{fig:2}(a). The fitted $\beta < 1$ indicates a sub-exponential decay, demonstrating that visual coupling persists farther than expected from simple contrast attenuation and sustains measurable intergroup coordination over separations exceeding the average nearest-neighbor distance, $d^{\text{NN}}=5.5\pm0.5$ cm (see SF Fig.~\ref{fig:d_nn}). An exponential form becomes consistent with the observations only at group separations well beyond this characteristic nearest-neighbor scale (see Fig.~\ref{fig:fit_interaction}).

A reasonable hypothesis to explain these findings is that sensory processing changes with scale: At short separations individual fish can resolve and respond to the motion of specific neighbors across the partition, producing a stronger, “pairwise‑like’’ contribution to intergroup coupling. By contrast, at larger separations (comparable to the group size) single individuals can no longer discriminate conspecifics in the distant subgroup; here the relevant signal is a coarse, group‑level motion cue and coupling is mediated by collective perception \cite{Davidson2021} rather than by identifiable pairwise interactions. As physical distance increases animals therefore shift from processing detailed, individual‑level information to relying on more abstract, ensemble‑level representations — an effect reminiscent of distinctions between peripersonal and extrapersonal perception in vision science.

Consistent with this interpretation, the spatial distribution of group centers of mass is similar for $d=0$ (single transparent wall) and $d=4$ cm, with higher occupancy near the partitions (see Fig.~\ref{fig:spatial_distribution}), whereas at $d =10$ cm we observe a contraction of the CM distribution across replicates, which can signal a transition in the dominant perceptual mode. Whether this change reflects an emergent sensitivity to remote collective motion (for example, as an adaptive mechanism to detect distant threats \cite{Li2024}) remains an open question that requires targeted experiments and is beyond the scope of the present study.

Together, these results indicate that schooling interactions are governed
primarily by visual coupling over the schools' effective sensory range. Our
findings refine the interaction structure of schooling systems: Although broadly
consistent with non‑metric or topological descriptions
\cite{balleriniInteractionRulingAnimal2008}, they reveal a graded mechanism in
which effective coupling is set by sensory‑mediated detectability and decays
smoothly with distance, rather than by a sharp interaction radius. Remarkably,
intergroup separations in our experiments exceed the typical nearest‑neighbor
spacing within each school, yet measurable coupling persists, demonstrating that
information transfer can operate beyond the scale of local cohesion.

\begin{figure}
    \centering
    \includegraphics[width=\linewidth]{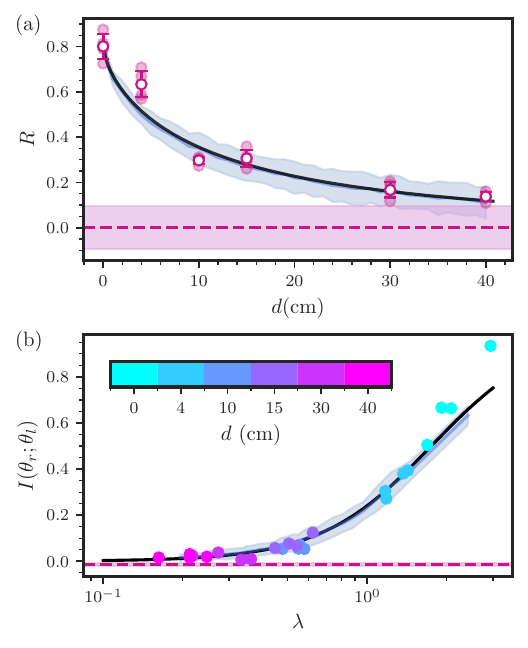}
    \caption{(a) Phase-locking order parameter $R=\langle \cos\varphi \rangle$ as a function of the gap distance. Solid markers denote the mean over experimental replicas, with error bars indicating the standard deviation across replicas; semi-transparent markers correspond to individual replicas. (b) Mutual information between $\theta_r$ and $\theta_l$ as a function of the inferred coupling $\lambda$, corrected using time-shift surrogates (see Appendix~\ref{sec:mutual_info}); marker colors indicate the gap distance. In both panels, the black curve shows the exact theoretical prediction obtained by numerical integration of Eq~\eqref{eq:R} and Eq.~\eqref{eqn:mutual_info} respectively, the blue curve the mean obtained from stochastic simulations of the Langevin model (shaded band: $\pm 2\sigma$), see Sec.~\ref{sec:simulation} for details, and the dashed line and shaded band the mean and standard deviation of the opaque-wall (uncoupled) condition $d\to\infty$.}
    \label{fig:3}
\end{figure}

\section{Phase locking and mutual information}

To quantify the transition from synchronized to desynchronized collective dynamics, we introduce a phase-locking order parameter defined as the first circular moment of the phase difference:
\begin{equation}
R = \langle \cos(\theta_r - \theta_l) \rangle = \langle \cos\varphi \rangle,
\label{eq:order_parameter}
\end{equation}
where $\langle \dots \rangle$ denotes a time and ensemble average and $R \in
[-1, 1]$ \cite{mardia1999directional}.
$R=1$ indicates perfect parallel synchronization ($\varphi=0$), $R=0$ symmetric switching between equally probable parallel and antiparallel states, and $R=-1$ persistent anti-synchronization, not observed here.  Fig.~\ref{fig:3}(a) shows that  $R(d)$ decreases smoothly from synchronized ($R\simeq0.8$) to desynchronized ($R\simeq0$) dynamics as the gap distance $d$ increases, in excellent agreement with the theoretical curve derived from our model, see Appendix~\ref{sec:sinchro_parameter}.

Beyond directional alignment, synchronization reflects statistical dependence and information sharing between the two collective degrees of freedom. To quantify this interaction, we evaluate the Mutual Information (MI) between school headings, $I(\theta_r; \theta_l)$. Remarkably, our analytical stationary distribution $P(\theta_r, \theta_l)$ yields a direct, parameter-free prediction for the MI across all separations once $\lambda(d)$ is determined.

Empirically, estimating MI from finite, auto-correlated trajectory time series introduces a positive baseline bias even in the uncoupled (opaque-wall) condition. We correct for this by implementing a time-shift procedure that preserves internal trajectory correlation times while destroying intergroup correlations (see Appendix~\ref{sec:mutual_info}). After subtracting this offset, the measured MI agrees quantitatively with our theoretical prediction across all tested gap separations [Fig.~\ref{fig:3}(b)]. This remarkable agreement confirms that a single distance-dependent coupling parameter $\lambda(d)$, extracted purely from phase-difference statistics, fully dictates both physical coordination and information transfer between spatially separated groups.

\section{Conclusions}

In summary, we have shown experimentally that spatially separated fish schools interacting primarily through visual cues synchronize their collective headings in a distance-dependent manner. Coarse-graining each school by its center-of-mass orientation revealed intrinsic bistable switching in isolation, which gives way to phase-locked synchronization when visual communication is permitted. Mapping these dynamics onto an analytically tractable model of two coupled bistable stochastic oscillators yields a single effective coupling parameter, $\lambda(d)$. Remarkably, this parameter, inferred solely from phase-difference statistics, quantitatively predicts both the degree of phase synchronization and the mutual information shared between schools across all spatial separations.

The spatial decay of the inferred coupling $\lambda(d)$ suggests a scale-dependent perceptual shift in intergroup communication. At short separations, fish can resolve and respond to specific neighbors across the partition, giving rise to stronger, pairwise-like interactions. At larger separations, individual resolution degrades, and coupling is mediated by coarse-grained, ensemble-level motion cues. These findings show that long-range, optical interactions can drive both macro-scale directional coherence and meaningful information exchange in biological collectives.

Methodologically, our approach combines controlled partitioned experiments, high-resolution multi-animal tracking, and stochastic theory to map complex behavioral observables to minimal physical parameters. Concretely, our results establish geometrically constrained fish schools as a physical realization of coupled bistable stochastic oscillators. Beyond schooling dynamics, this  framework offers a quantitative foundation for studying sensory-mediated communication in active matter, with natural extensions to multimodal sensing, fission-fusion dynamics, and the design of bio-inspired distributed robotic systems.

\begin{acknowledgments}
We acknowledge financial support from projects PID2022-137505NB-C21 and
PID2022-137505NB-C22, funded by MICIU/AEI/10.13039/501100011033, and by the
European Regional Development Fund (ERDF): ``A way of making
Europe''. R.P.-S. acknowledges support from the Acad\`emia d'Excel$\cdot$l\`encia
programme, funded by the Generalitat de Catalunya.
\end{acknowledgments}

\appendix

\section{Experimental Methods}\label{sec:experimental_methods}

\subsection{Model System and Ethical Statement}

All experiments were conducted at the Scientific and Technological Centers UB (CCiTUB), University of Barcelona (Spain), under the review and approval of the University of Barcelona Ethics Committee (project number 119/18). We utilized black neon tetra (\textit{Hyphessobrycon herbertaxelrodi}), a small freshwater species with a mean body length of approximately $2.5$cm. This species is an ideal experimental realization for low-dimensional active matter studies because it naturally forms highly coherent, strongly polarized, and effectively planar schools~\cite{gimenoDifferencesShoalingBehavior2016}.

\subsection{Apparatus and Experimental Matrix}

Experiments were conducted in a $100 \times 100 \times 40$~cm$^3$ square tank filled with water to a depth of 7 cm. Because these fish predominantly swim near the surface, collective motion was effectively two-dimensional.

To systematically tune long-range intergroup coupling, the tank was partitioned into three water-filled sections using transparent or opaque walls (see Fig.~\ref{fig:1}(a) in the main text). The experimental matrix explored three structural regimes:
\begin{itemize}
    \item Intermediate gaps: Two transparent partitions set the separation distance to $d = 4, 10, 15, 30,$ and $40$ cm), with one school of $N=10$ fish in each outer compartment.
    \item Maximal visual coupling ($d = 0$ cm): A single central transparent wall separated the groups, with $N = 20$ fish per side.
    \item Zero visual coupling ($d=\infty)$: A central opaque wall blocked visual contact, also with $N = 20$ fish per side.
\end{itemize}
The larger group size at $d=0$ and $d=\infty$ reduced finite-size fluctuations, providing stable asymptotic baselines for the intermediate-gap conditions.

For each value of $d$, we evaluated four independent $10$-minute replicates, obtained from non-overlapping segments of longer continuous recordings, with durations chosen to capture multiple stochastic switching events between the antiparallel orientation states.

\subsection{Data Acquisition and Digitization}

Video sequences were recorded at $50$ fps with a resolution of $5312 \times 2988$ pixels using a GoPro Hero 11 Black digital camera mounted $1$ m directly above the center of the tank. The outer boundary of the tank spanned $L = 2730$ pixels in the recordings, defining the scaling factor for coordinate conversion. To eliminate lens distortion, a spatial perspective correction was applied to the tank region during image preprocessing~\cite{puySelectiveSocialInteractions2024}.

Individual trajectories were extracted using \texttt{idtracker.ai}~\cite{romero-ferreroIdtrackerAiTracking2019}. Occasional occlusions causing invalid position data were corrected using the Validator tool in \texttt{idtracker.ai} v5. To mitigate high-frequency experimental noise, the raw time series were smoothed with a 1D Gaussian filter ($\sigma = 2$ frames, truncated at $5\sigma$) implemented via \texttt{scipy.ndimage.gaussian\_filter1d}~\cite{virtanenSciPyFundamentalAlgorithms2020}. Instantaneous velocities and accelerations were obtained by differentiating the smoothed trajectories using the first and second derivatives of the Gaussian kernel, respectively~\cite{nixonFeatureExtractionImage2010}.

\section{Maximum Likelihood Parameter Inference and simulation protocal}\label{sec:mle_fitting}

To quantitatively compare experimental trajectories with our Fokker--Planck framework, we parameterize the stationary  phase difference distributions $P(\varphi)$ using the dimensionless parameters $\kappa = K/D$ (bistability strength) and $\lambda = J/D$ (coupling strength). Rather than performing least-squares fits on binned histogram densities, parameters are inferred via Maximum Likelihood Estimation (MLE) directly on the unbinned experimental time series, eliminating bin-size dependencies and edge artifacts~\cite{gregory2010bayesian}.

\subsection{Parameter Inference}

The stationary distribution of the phase difference $\varphi$ derived from the Fokker-Planck equation is given by:
\begin{equation}
p(\varphi \mid \kappa, \lambda) = \frac{1}{Z_\varphi(\kappa, \lambda)} \exp(\lambda \cos\varphi) I_0(2\kappa \cos\varphi),
\end{equation}
where $Z_\varphi(\kappa, \lambda)$ is the normalization partition function evaluated numerically over $\varphi \in [-\pi, \pi]$. Given empirical observations $\{\varphi_i\}_{i=1}^N$ at a given gap $d$, the log-likelihood reads:
\begin{equation}
\mathcal{L}(\kappa, \lambda) = \sum_{i=1}^N \ln p(\varphi_i \mid \kappa, \lambda) = \sum_{i=1}^N \left[ \lambda \cos\varphi_i + \ln I_0(2\kappa \cos\varphi_i) \right] - N\ln Z_\varphi(\kappa, \lambda).
\end{equation}
The distance-dependent optimal  $\kappa(d)$ and coupling strength $\lambda(d)$ are extracted by minimizing $-\mathcal{L}(\kappa, \lambda)$.

\subsection{Simulations protocol}\label{sec:simulation}

The parameters $\kappa=K/D$ and $\lambda=J/D$  quantify the relative well depth and interaction strength, respectively. Since the stationary distributions depend only on these ratios, we set $D=1$ without loss of generality in simulations. For each condition, $\kappa$ was fixed to its mean value across realizations, and $\lambda$ was obtained from Eq.~\eqref{eq:stretched_exp} using the fitted parameters.

The stochastic dynamics were simulated by numerically integrating the Langevin equations over $10^5$ time steps using the stochastic Runge--Kutta solver \texttt{itoSRI2} from the \texttt{sdeint} package \cite{sdeint}. The first $1\times10^4$ integration steps were discarded as a burn-in period to ensure that the system had reached its stationary regime before collecting statistics.

For each parameter set, 50 independent realizations were generated from random initial conditions. The reported averages correspond to the mean over these realizations, while the error bars shown in Fig.~\ref{fig:3} represent two standard deviation across the independent simulations.

\section{Phase locking order parameter}\label{sec:sinchro_parameter}

\begin{figure}
    \centering
    \includegraphics[width=\linewidth]{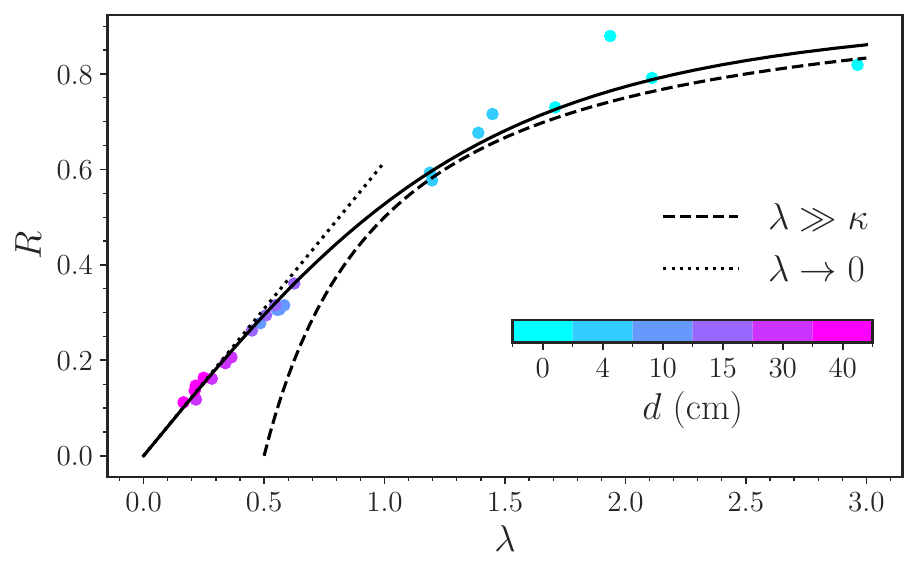}
    \caption{
Phase-locking order parameter $R$ as a function of the coupling parameter $\lambda$. Solid points denote experimental data, the solid line the exact solution, the dotted line the weak-coupling approximation, and the dashed line the strong-coupling approximation.
}
\label{fig:order_param_approx}
\end{figure}

The phase locking order parameter is defined  as $R=\langle\cos\varphi\rangle$ and can be evaluated exactly as
\begin{equation}
R=
\frac{
\int_{-\pi}^{\pi}
\cos\varphi \;
e^{\lambda\cos\varphi}
I_0\left(2\kappa\cos\varphi\right)
d\varphi
}{
\int_{-\pi}^{\pi}
e^{\lambda\cos\varphi}
I_0\left(2\kappa\cos\varphi\right)
d\varphi
}.
\label{eq:R}
\end{equation}
While this expression has no simple closed form, limiting regimes can be obtained
analytically. For $\lambda=0$, symmetry gives $R=0$, corresponding to the
unsynchronized state. For weak coupling, expanding $e^{\lambda\cos\varphi}\approx 1+\lambda\cos\varphi$ yields
\begin{equation}
  R \approx \lambda
    \frac{\displaystyle\int_{-\pi}^{\pi}\cos^2 \;\varphi I_0\big(2\kappa\cos\varphi\big)d\varphi}
    {\displaystyle\int_{-\pi}^{\pi}I_0\big(2\kappa\cos\varphi\big)d\varphi}
    \equiv \lambda\chi(\kappa),
\end{equation}
where $\chi(\kappa)$ is the linear susceptibility. Conversely, for strong coupling
$\lambda\gg\kappa$,
\begin{equation}
R\simeq \frac{I_1(\lambda)}{I_0(\lambda)}
\simeq
1-\frac{1}{2\lambda}
+O\!\left(\lambda^{-2}\right).
\end{equation}
Fig.~\ref{fig:order_param_approx} summarizes these results.

\section{Mutual Information Analysis}\label{sec:mutual_info}

Mutual information (MI) quantifies the statistical dependence between two random variables, or the reduction in uncertainty about one variable obtained by observing the other \cite{cover2006elements}. It has previously been used to characterize synchronization between oscillators, including in electroencephalographic signals \cite{PhysRevE.65.041903}. We therefore quantify information sharing between the two groups by computing the MI between their heading time series $\theta_r$ and $\theta_l$,
\begin{equation}\label{eqn:mutual_info}
    {\displaystyle \operatorname {I}(\theta_r; \theta_l)=\iint P(\theta_r', \theta_l') \log {\left({\frac {P(\theta_r', \theta_l')}{P(\theta_r')\,P(\theta_l')}}\right)}\;d\theta_r'\,d\theta_l'},
\end{equation}
where $P(\theta_r,\theta_l)$ is the joint probability density of the headings and $P(\theta_r)$ and $P(\theta_l)$ are the corresponding marginal densities. From the joint distribution in the variables $\varphi$ and $\psi$, Eq.~\eqref{eq:joint}, the heading distribution follows from the change of variables $\theta_r=(\psi+\varphi)/2$ and $\theta_l=(\psi-\varphi)/2$:
\begin{equation}
    P(\theta_r,\theta_l)
    =
    \frac{1}{Z}
    \exp\left[
    \kappa\big(\cos 2\theta_r+\cos 2\theta_l\big)
    +\lambda\cos(\theta_r-\theta_l)
    \right].
\end{equation}
The corresponding marginal distribution is
\begin{equation}
\begin{split}
    P(\theta_i)
    &=
    \frac{2\pi}{Z}
    \exp\left(\kappa\cos 2\theta_i\right)
    \Biggl[
    I_0(\kappa)I_0(\lambda)\\
    &+
    2\sum_{n=1}^{\infty}
    I_n(\kappa)I_{2n}(\lambda)\cos(2n\theta_i)
    \Biggr],
\end{split}
\end{equation}
where $I_n$ is the modified Bessel function of the first kind of order $n$~\cite{abramovitz1972}. Inserting these distributions into Eq.~\eqref{eqn:mutual_info} allows the MI to be computed by numerical integration and compared with simulations, as shown in Fig.~\ref{fig:3}b.

Mutual information estimated directly from experimental data using {\tt mutual\_info\_regression} from Scikit-learn~\cite{scikit-learn} yielded a nonzero value even in the opaque, uncoupled condition, $I^{\rm opaque}(\theta_r,\theta_l)=0.46\pm0.07$. This finite value is inconsistent with the theoretical expectation for $\lambda=0$, where $P(\theta_r,\theta_l)=P(\theta_r)P(\theta_l)$ and hence
\begin{equation}
\log\left[
\frac{P(\theta_r',\theta_l')}
{P(\theta_r')P(\theta_l')}
\right]=0 .
\end{equation}
The discrepancy is attributable to the known positive bias of mutual-information estimators for finite, temporally correlated data sets, for which the effective number of independent samples is reduced~\cite{PhysRevE.71.066208}. We estimated this bias using circular-shift surrogates: one trajectory was shifted
relative to the other by a random lag exceeding the autocorrelation time, estimated
from the first minimum of $I(X;X_\tau)$ as $\tau_{\rm lag}\simeq2.28$~s. This
preserves each trajectory's marginal statistics and temporal correlations while
destroying temporal correspondence between trajectories~\cite{PhysRevE.65.041903,PhysRevE.66.041904}.
The corrected mutual information was computed as
\begin{equation}
I = I_{\rm observed}-\langle I_{\rm surrogate}\rangle,
\end{equation}
where $\langle I_{\rm surrogate}\rangle$ is the average over 500 independently
shifted surrogate realizations.

\section{Supplemental figures}
\vspace{-1cm}
\begin{figure*}[h]
    \centering
    \includegraphics[width=0.8\textwidth]{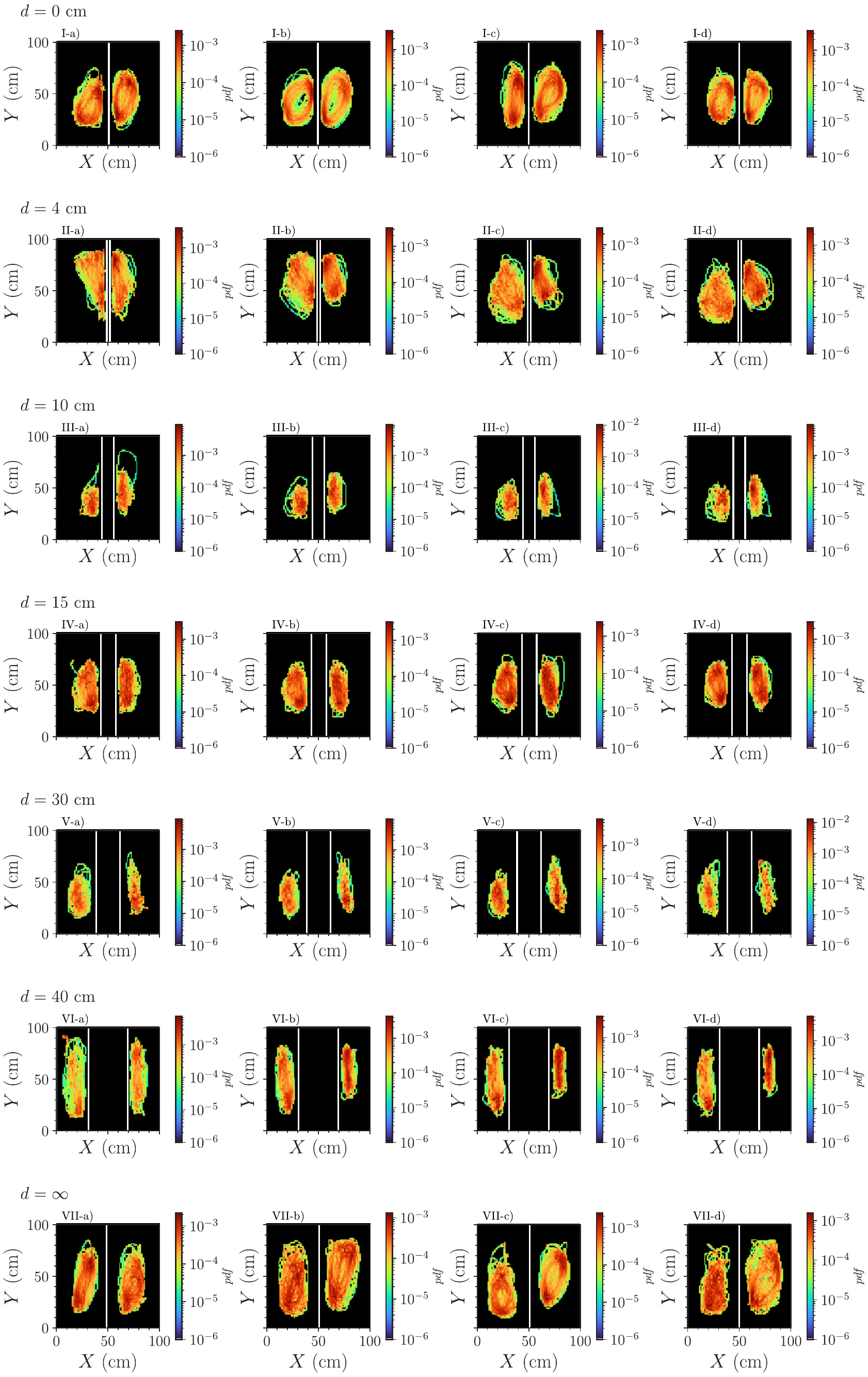}
\caption{Empirical steady-state probability density $P(X,Y)$ of the schools' center of mass, estimated from normalized two-dimensional spatial histograms. Boundary confinement concentrates the density along the long-wall axis, yielding effectively one-dimensional spatial dynamics. Rows I--VII correspond to $d=0,4,10,15,30,40$~cm, and $d=\infty$, respectively; columns a)--d) denote the four replicate trials. White lines indicate the wall positions.
}
    \label{fig:spatial_distribution}
\end{figure*}

\begin{figure*}
    \centering
    \includegraphics[width=\textwidth]{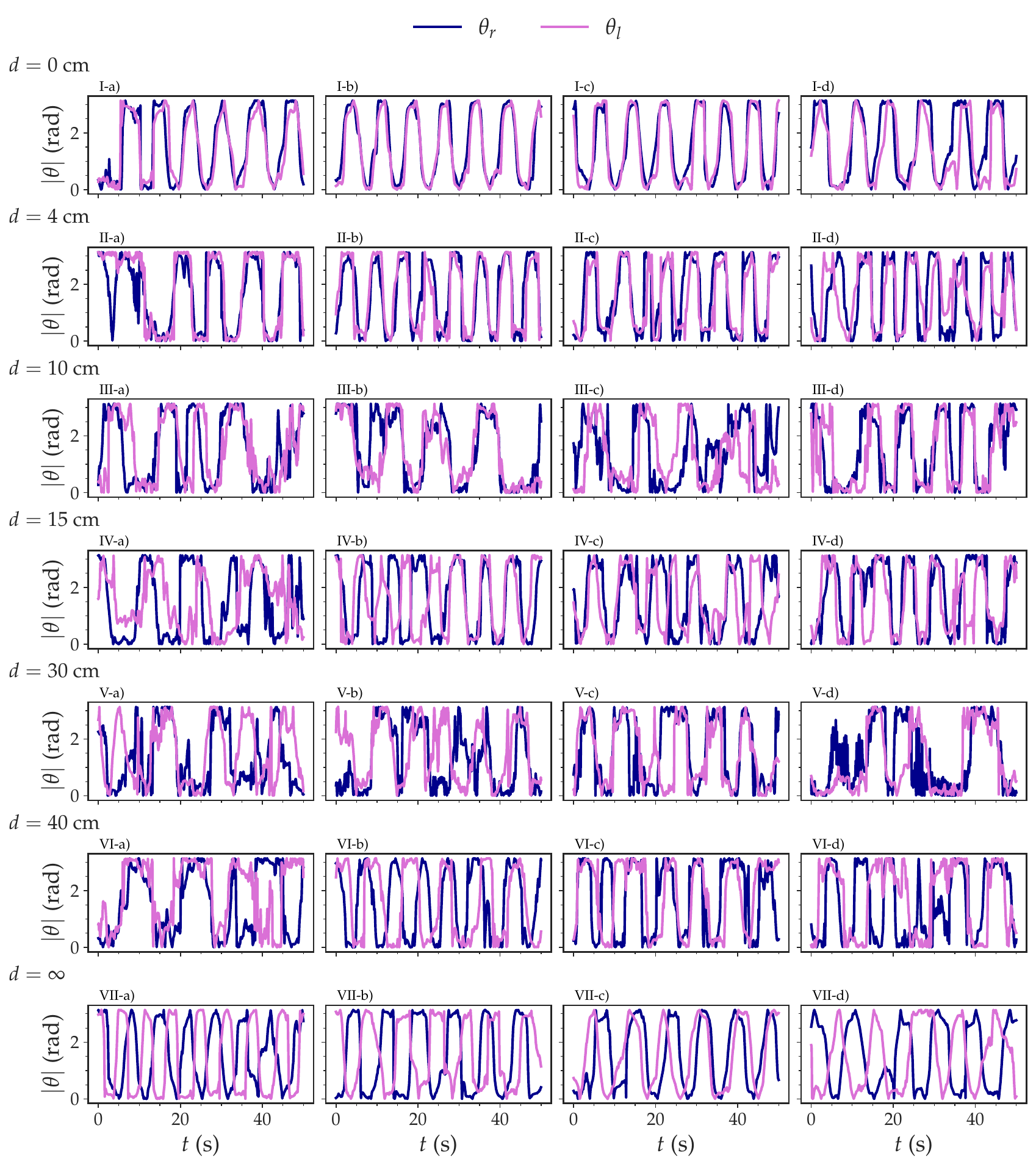}
    \caption{Time series of the center‑of‑mass orientation $\theta$ across all separations and replicas. Rows I-VII correspond to $d=0,4,10,15,30,40$ cm, and $d=\infty$, respectively; columns a)--d) show the different trials.}
    \label{fig:bistability}
\end{figure*}

\begin{figure*}
    \centering
    \includegraphics[width=\textwidth]{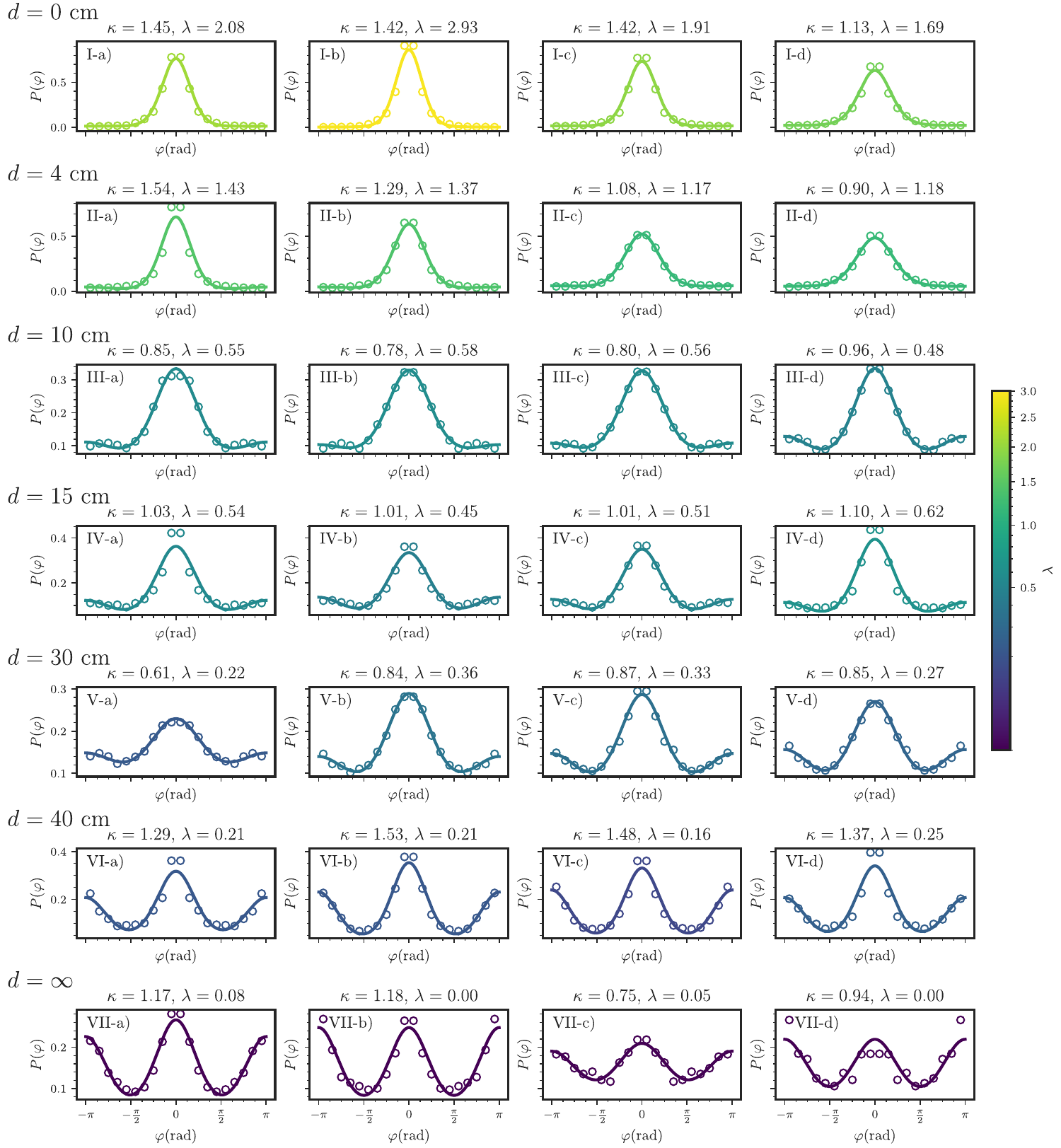}
    \caption{
Phase-difference distributions $P(\varphi)$ between the two subgroups for all separation distances and replicate trials. Rows I--VII correspond to $d=0,4,10,15,30,40$~cm, and $d=\infty$, respectively; columns a)--d) denote the four replicate trials. Values of $\kappa$ and $\lambda$ displayed in each panel are obtained by maximum-likelihood fitting. Data were symmetrized under $\varphi \mapsto -\varphi$ to enforce the expected reflection symmetry.
}
    \label{fig:phi}
\end{figure*}

\begin{figure}
    \centering
    \includegraphics[width=\linewidth]{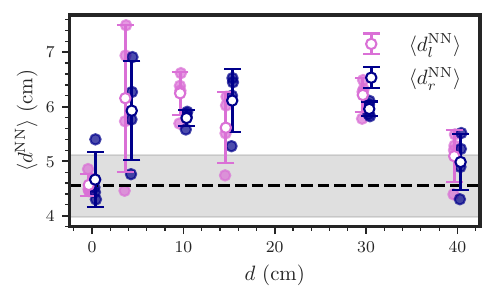}
    \caption{Mean nearest-neighbour distance $\langle d^{\rm NN}\rangle $ within each subgroup as a function of intergroup distance $d$. Colors distinguish the right and left groups. Small filled markers show individual realizations, horizontally jittered for clarity, whereas large open markers and error bars indicate the mean and standard deviation across realizations. The dashed line and shaded band denote the mean and standard deviation for the $d\to \infty$ condition. Across experiments, the mean nearest-neighbor distance is $ d^\text{NN}=5.5\pm 0.5\text{ cm}$.}
    \label{fig:d_nn}
\end{figure}

\begin{figure}
    \centering
    \includegraphics[width=\linewidth]{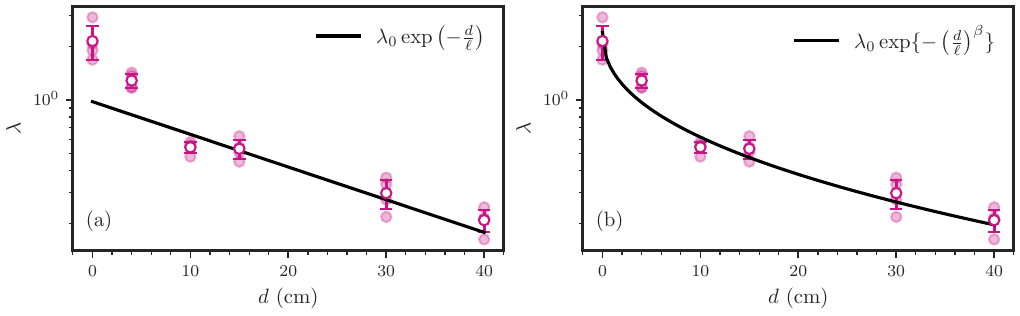}
    \caption{Distance dependence of the average coupling parameter $\lambda(d)$.
(a) Exponential fit to the experimental mean values of $\lambda$, plotted on a logarithmic scale, yielding $\lambda_0 = 1.0 \pm 0.1$ and $\ell = 23 \pm 4$~cm. The fit is performed over the full dataset using the standard deviations of the mean values as uncertainties and underestimates the coupling at short distances, notably $d=0$ and $4$~cm.
(b) Stretched-exponential fit to the experimental mean values of $\lambda$, yielding $\lambda_0 = 2.41 \pm 0.08$, $\ell = 4.9 \pm 0.7$~cm, and $\beta = 0.44 \pm 0.04$. Central values are obtained from the full-data fit, while uncertainties are estimated by a jackknife procedure in which one replicate is omitted independently at each distance and the fit is repeated over all leave-one-out combinations.
}

    \label{fig:fit_interaction}
\end{figure}

\clearpage

\section{Supplemental videos}

\begin{video}[h]
\caption{Representative experiment for a wall separation distance of $d=4$ cm with $N=10$ fish in each compartment. Fish trajectories over the previous $1.2$ s are shown, and large arrows indicate the center-of-mass velocity of each group. The left panels display the instantaneous heading direction and the absolute heading angle of each group. The video shows synchronization of the two groups through the transparent walls and the bistable switching of the heading angle arising from their motion along the long axis of the compartments.}
\label{vid:video1}
\end{video}

\end{document}